\documentstyle[epsf,rotate]{l-aa}
\def\ha{H$\alpha$ }

\def\h0{{\rm H_0}}

\begin{document}

\thesaurus{07 (08.05.2;08.03.4;08.14.1;08.09.2 V635 Cas; 08.09.2 Hiltner 102)}

\title{Optical Spectroscopy of V635 Cassiopeiae/4U~0115+63}

   \author{ S. J. Unger \inst{1}
   \and P. Roche \inst{2}
   \and I. Negueruela \inst{3}
   \and F. A. Ringwald \inst{4}
   \and C. Lloyd    \inst{5}
   \and M. J. Coe \inst{6}	  
          }

\offprints{sju@ipac.caltech.edu}

\institute{IPAC, M.S. 100-22, California Institute of Technology, Pasadena, CA 91125, U.S.A. 
\and Astronomy Centre, CPES, Sussex University, Falmer, Brighton, BN1 9QJ, U.K.\and Astrophysics Group, Liverpool John Moores University, Byrom St., Liverpool, L3 3AF, U.K.  
\and  Dept. of Astronomy \& Astrophysics, Pennsylvania State University, 525 Davey Laboratory, University Park, PA 16802, U.S.A. 
\and Rutherford Appleton Laboratory, Chilton, Oxon, OX11 0QX, U.K. 
\and Physics Dept. Southampton University, Southampton SO17 1BJ, U.K}

   \date{Accepted 1998 January 21}

   \maketitle

   \begin{abstract}

V635 Cas is the  optical counterpart of the X-ray binary system 4U~0115$+$63. 
It was previously tentatively identified as a Be star based on its optical 
colours and the presence of \ha emission. Our observations 
indicate that it is an O9e star. This is the first direct determination of 
this star's optical
spectral type. The presence of a hotter companion star
may in part explain the large temporal variation observed in this system. 

Extreme variability was observed in 1992 February when both the \ha and 
a series of Paschen lines changed from emission to
absorption. This was interpreted as a disk-loss event and it is the first 
time that it has been observed in this system.
We use far red spectra of V635 Cas to
probe the circumstellar disk, discussing the various line formation regions.
The lines observed are consistent with a late type Oe star. 

The flux standard Hiltner 102 was also observed. Although it is classified 
as a B0 III star, we re-classify it as a O9.7 II star 
with a slight nitrogen enhancement.


\end{abstract}

\section{Introduction}

The system V635 Cas/4U~0115+63 is a Be X-ray binary star system (BeXRB)
with a 3.61s spin period and a 24.3 day orbital period (Cominsky et al.
1978; Rappaport et al. 1978).  The optical counterpart, V635 Cas, was
tentatively classified as a early type Be star based on its optical
colours and the presence of variable H$\alpha$ and occasional H$\beta$
emission (Johns et al. 1978;  Kriss et al. 1983; Hutchings \& Crampton
1981). The temporal evolution of V635 Cas is very different from that of
other BeXRBs. The X-ray outbursts from the neutron star are varied in
strength but typically last for a month. The associated optical and
infrared activity is far more prolonged, lasting typically $\sim$6 months. 
Unlike many BeXRBs the peak in the X-ray flux is not centred on periastron
passage.  This suggests that it is the episodic equatorial mass loss from
the companion star which is the trigger for each outburst.  Mendelson \&
Mazeh (1991) concluded that X-ray outbursts occur when the optical
outburst is relatively long ($\sim$200 days) and strong ($\sim$1 mag).
X-ray emission during weaker optical outbursts is not seen due to
centrifugal inhibition of matter and the propeller effect (Kriss et al.
1983; Mendelson \& Mazeh 1991). 

Negueruela et al. (1997, Paper I) discuss the May-June 1994 X-ray outburst
in the context of long term observations of V635 Cas. We conclude that the
large variations in optical luminosity originate in the Be circumstellar
envelope and not an accretion disk around the neutron star. The orbit of
the neutron star is relatively close to the companion and its
gravitational pull may play an important role in the evolution of the
circumstellar disk. 

In this paper we present two data sets that can be used to constrain the
physical and geometric models for the circumstellar disk. The companion's
spectral type is a critical parameter for modelling the disk. However this
is hard to derive as many of the companion stars in the BeXRBs are faint
and the photospheric lines are often filled in with disk emission. Many of
the systems have a spectral classification based on optical colours. In
Sect. 3.1 we present the first blue spectra of V635 Cas with sufficient
signal to noise to derive a spectral class. 

The disks in Oe/Be stars are highly variable. We need a database of high
quality data in order to model the evolution of the disk size, temperature
and density. In Sect. 3.2 we present the first far red optical spectra
obtained for this system.

\begin{figure}
\rotate[l]{
\epsfxsize=68mm
\epsfbox[100 0 530 630]{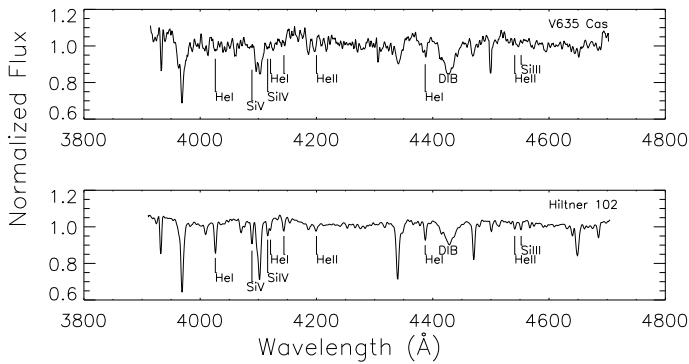}}
\caption[]{$\lambda\lambda$ 3900 -- 4700 \AA\ Spectra: V635 Cas
(upper) and Hiltner 102 (lower)}
\end{figure}

\section{Observations}

\subsection{Blue Spectra}

The spectra were obtained in service mode with the William Herschel
Telescope, La Palma using ISIS with the TEK CCD camera and the R600B (0.79
\AA\ pixel$^{-1}$ dispersion, 1.6 \AA\ pixel$^{-1}$ resolution, range
$\lambda\lambda$ 6350-6750 \AA) and R1200R (0.41 \AA\ pixel$^{-1}$
dispersion, 0.8 \AA\ pixel$^{-1}$ resolution, range $\lambda\lambda $
3900-4700 \AA) gratings (Carter et al. 1993). The simultaneous
observations of \ha allowed us to assess any emission component that may
be present in the bluer Balmer lines. \ha was in emission at this time,
see Paper I. 

V635 Cas was observed over two nights (900s and 2$\times$1500s on 1993
December 18 and 2$\times$1000s on 1993 December 19). The flux standard
Hiltner 102 was also observed (300s on 1993 December 18). The seeing was
poor on 1993 December 18. The data were reduced using the {\small FIGARO}
and {\small DIPSO} packages (Shortridge \& Meyerdicks 1996; Howarth 1996). 

Figure 1 is a plot of V635 Cas (upper) and Hiltner 102 (lower) obtained on
1993 December 19 and 18, respectively.  The data have been smoothed with a
Gaussian ($\sigma$ = 2, width = 5 pixels, 1 pixel = 0.8 \AA).  To
normalise the spectra we fitted a third order polynomial to the continuum
and then divided the spectra by this fit.

\subsection{Red Spectra}

The far red spectra ($\lambda\lambda$ 6400--8900 \AA, 5 \AA\ pixel$^{-1}$
dispersion, 11 \AA\ resolution) were obtained using the all-transmission
Mark IIIa spectrograph on the 1.3m McGraw-Hill telescope at
Michigan-Dartmouth-MIT Observatory, Kitt Peak, Arizona. The detector was a
TI-4849 CCD, inside the BRICC camera (Luppino 1989), except on 1991
October 27, when a Thomson chip was used. All spectra were taken through a
2.2$^{\prime\prime}$ slit, with exposure times of 1800 s.  There was
cirrus on 1991 October 20. 

Spectra of hot stars were taken to map the telluric atmospheric bands in
this part of the spectrum, which were removed using the methods of Wade \&
Horne (1988). There is a wrinkle in the spectra at 7600 \AA\ due to the
removal of the A-band which is the strongest atmospheric feature. There is
also an atmospheric absorption band in the region P11 to P7. Each time a
spectrum of V635 Cas was taken, one spectrum each of two flux standards
were also taken. These were G191B2B (Oke 1974) and HD 19445 (Oke \& Gunn
1983). Comparing the individual spectra of these standards, taken at
different epochs, we found the relative fluxes in the spectra to be
consistent to within a few percent. Errors in relative fluxes due to
losses from an unrotated slit were therefore not serious, unsurprising
since these spectra are all so red (Filippenko 1982). Absolute fluxes were
another matter: we estimate from the same spectra that the absolute flux
levels should not be trusted to within 30\%. Although this will effect
flux measurements of the lines, it will not affect equivalent width or
velocity measurements. The instrumental uncertainties do not account for
the upturn in the spectrum on 1991 October 27 and we believe this to be
real. 

The $\lambda\lambda$ 6400 -- 7500 \AA\ and $\lambda\lambda$ 7600 -- 8900
\AA\ spectra are given in Fig. 2.  The 1991 October 20 spectrum has been
normalised to the continuum value at 7264 \AA\ as the night was not
photometric. 

Tables 2 and 3 list the dereddened line fluxes.  The spectra were
dereddened assuming a standard Galactic extinction law (Rieke \& Lebofsky
1985; Howarth 1983) and $E(B - V) = 1.5$ (Hutchings \& Crampton 1981).

\begin{table}
\begin{center}
\caption{The Observed Lines in The Far Red CCD Spectra}
\vspace*{0.3cm}
\begin{tabular}{l l l l}
&&&\\
{Element}&{$\lambda_{lab}$}&{$\lambda_{obs}$}&{Notes}\\
&{\ \ \ \AA}&{\  \  \ \AA}&\\
&&&\\
Fe\,{\sc ii}&6473.9&6481&\\
Fe\,{\sc ii}&6506.3&6506&\\
H$\alpha$&6562.8&6562&\\
He\,{\sc i}&6678.15&6678&\\
He\,{\sc i}&7065.19&7064&\\
C\,{\sc ii}&7231&7235&\\
&7236&&\\
He\,{\sc i}&7281.35&7280&\\
K\,{\sc i} &7664.90&7662& \\
K\,{\sc i}&7698.96&7698&\\
O\,{\sc i}&7771.96&7769&\\
&7774.18&&\\
&7775.40&&\\
P19&8413.32&8411&\\
O\,{\sc i}&8426.16&&blended with P18 \\
P18&8437.96&8436 &\\
P17&8467.26&8464 &\\
Ca\,{\sc ii}&8498.02&&blended with P16 \\
P16&8502.49&8499&\\
Ca\,{\sc ii}&8542.09&&blended with P15 \\
P15&8545.39&8542&\\
P14&8598.39&8596 &\\
Ca\,{\sc ii}&8662.14&&blended with P13 \\
P13&8665.02&8662 &\\
N\,{\sc i}&8629.2&&lines blended\\
&8680--6&&\\
& 8703&&\\
&8712--9&&\\ 
P12&8750.5&8748 &\\
P11&8862.79&8862&\\              
\end{tabular}
\end{center}
Line identification from Meinel et al. (1975); the line centres can vary
due to blending; the presence of O\,{\sc i} 8446 \AA\ and Ca\,{\sc ii}
triplet can only be inferred by an increase in the relative strengths of
the Paschen lines.  The N\,{\sc I} and K\,{\sc I} are only suspected to be
present, see text. 
\end{table}
\normalsize
%\clearpage
\nopagebreak
\small
\begin{table}
\begin{center} 
\caption{Line Parameters \ha $\lambda$ 6563 \AA\ -- O\,{\sc i} $\lambda$ 
7772 \AA} 
\vspace*{0.3cm}
\begin{tabular}{lcccc}
{TJD$^{\dagger}$}&{Date}&{Line}&{EW}&{Flux$^{\dagger\dagger}$}\\
%&&&&{$\times10^{-13}$}&\\
&&{\AA}&{\AA}&\\
&&&&\\
8546&91 Oct 16&H$\alpha$&--15.2&13.6\\
8550&91 Oct 20&6563&--17.8&--\\
8557&91 Oct 27&&--17.5&22.8\\
8672&92 Feb 19&&+1.2&1.54 \\
&&&&\\
8546&91 Oct 16&He~I&--2.3&1.98\\
8550&91 Oct 20&6678&--&--\\
8557&91 Oct 27&&--1.6&2.03 \\
8672&92 Feb 19&&+0.46&0.57\\
&&&&\\
8546&91 Oct 16&He~I&--2.1&1.60\\
8550&91 Oct 20&7065&--&--\\
8557&91 Oct 27&&--2.3&2.68\\
8672&92 Feb 19&&+0.47&0.48 \\
&&&&\\
8546&91 Oct 16&He~I&--1.5&1.07\\
8550&91 Oct 20&7281&--1.3&--\\
8557&91 Oct 27&&--0.4&0.46 \\
8672&92 Feb 19&&+1.2&1.1 \\
%&&&&&\\
&&&&\\
8546&91 Oct 16&K~I&+0.5&0.33\\
8550&91 Oct 20&7665&+1.4&--\\
8557&91 Oct 27&&+0.5&0.50 \\
8672&92 Feb 19&&+0.6 &0.43  \\
&&&&\\
8546&91 Oct 16&K~I&+0.9&0.50 \\
8550&91 Oct 20&7699&+0.9&--\\
8557&91 Oct 27&&+1.0&0.89\\
8672&92 Feb 19&&+1.0&0.72 \\
&&&&\\
8546&91 Oct 16&O~I&--0.5&0.28 \\
8550&91 Oct 20&7772&--1.6&--\\
8557&91 Oct 27&&--&--\\
8672&92 Feb 19&&--&-- \\
%&&&&\\
\end{tabular}
\end{center}
$^{\dagger}$ TJD =  JD $-$ 2440000 ;
$^{\dagger\dagger}$ Flux $\times10^{-13}$erg s$^{-1}$cm$^{-2}$
The line fluxes have been dereddened - see text.
\end{table}
\normalsize
%\clearpage

\nopagebreak
\small
\begin{table}
\begin{center} 
\caption{Line Parameters P19 $\lambda$ 8413 \AA--P11 $\lambda$ 8863 \AA}
\vspace*{0.3cm}
\label{ewll}
\begin{tabular}{lcccc}
{TJD$^{\dagger}$}&{Date}&{Line}&{EW}&{Flux$^{\dagger\dagger}$}\\
%{(2440000)}&&&&{$\times10^{-13}$}\\
&&{\AA}&{\AA}&\\
&&&&\\
8546&91 Oct 16&P19&--0.6&0.24 \\
8550&91 Oct 20&8413&--1.2&-- \\
8557&91 Oct 27&&-- & --\\
8672&92 Feb 19&&--&--\\
&&&&\\
8546&91 Oct 16&P18$^{*}$&--1.8&0.68\\
8550&91 Oct 20&8438&--3.2&-- \\
8557&91 Oct 27&&--1.3&1.03 \\
8672&92 Feb 19&&&  \\
&&&&\\
8546&91 Oct 16&P17&--2.3&0.83\\
8550&91 Oct 20&8467&--2.1&--\\
8557&91 Oct 27&&--1.7&1.3\\
8672&92 Feb 19&&--&--  \\
&&&&\\
8546&91 Oct 16&P16$^{*}$&--3.4&1.18 \\
8550&91 Oct 20&8502&--4.2&-- \\
8557&91 Oct 27&&--3.4&2.52\\
8672&92 Feb 19&&-&-  \\
&&&&\\
8546&91 Oct 16&P15$^{*}$&--4.6&1.55\\
8550&91 Oct 20&8545&--5.3&--\\
8557&91 Oct 27&&--3.8&2.71\\
8672&92 Feb 19&&--& --  \\
&&&&\\
8546&91 Oct 16&P14&--6.4&2.01\\
8550&91 Oct 20&8598&--6.4&--\\
8557&91 Oct 27&&--5.23&3.66\\
8672&92 Feb 19&&--& -- \\
&&&&\\
8546&91 Oct 16&P13$^{*}$&--6.1&1.87\\
8550&91 Oct 20&8665&--8.7&--\\
8557&91 Oct 27&&--4.4&3.07\\
8672&92 Feb 19&&+0.8&0.43\\
&&&&\\
8546&91 Oct 16&P12&--5.3&1.57\\
8550&91 Oct 20&8750&--5.8&-- \\
8557&91 Oct 27&&--6.2&4.25\\
8672&92 Feb 19&&+0.6&2.83\\
&&&&\\
8546&91 Oct 16&P11&--4.5&1.21\\
8550&91 Oct 20&8863&--2.9&--\\
8557&91 Oct 27&&--3.8&2.66\\
8672&92 Feb 19&&+2.9&1.44 \\
\end{tabular}
\end{center}
$^{\dagger}$ TJD = JD $-$ 2440000;
$^{\dagger\dagger}$ Flux $\times10^{-13}$erg s$^{-1}$cm$^{-2}$
The line fluxes have been dereddened - see text.
$^{*}$ blended, see Table 1.
\end{table}
\normalsize
%\clearpage

\section{Discussion}

\subsection{Spectral Classification}

The spectral class of V635 Cas was determined by comparison with Hiltner
102 and the standards published by Walborn \& Fitzpatrick (1990).  The
Walborn (1971) scheme hinges on the ratios of neutral and singly-ionised
helium and the first three ions of silicon. 

The comparison with the standard star Hiltner 102 which is identified as a
B0 III in the Simbad database led us to reclassify Hiltner 102 as a O9.7
II star. The spectral classification was based on the He\,{\sc ii}
$\lambda$ 4541 \AA / He\,{\sc i} $\lambda$ 4387 \AA\ and the He\,{\sc ii}
$\lambda $4200 \AA / He\,{\sc i} $\lambda$ 4144 \AA\ ratios. For an O 9.7
star the strength of the He\,{\sc ii} $\lambda$ 4541 \AA\ $\sim$ Si\,{\sc
iii} $\lambda $4552 \AA. The luminosity class was determined from the
Si\,{\sc v} $\lambda $4089 \AA / He\,{\sc i} $\lambda\lambda$ 4026, 4121,
4144 \AA\ and the Si\,{\sc iv} $\lambda$ 4116 \AA / He\,{\sc i} $\lambda$
4121 \AA\ ratios. Hiltner 102 may have a slight nitrogen enhancement. 

V635 Cas is harder to classify due to two factors. It is fainter than
Hiltner 102 ($V = 15.5$ vs. Hiltner 102, $V = 10.42$) and the disk
emission causes the filling in of the bluer singly ionized hydrogen and
helium lines. For example the filling in is evident in H$\gamma$ $\lambda$
4340 \AA. In addition the He{\sc ii} lines appear to be filled in. The
strong He\,{\sc ii} $\lambda$ 4200 \AA\ absorption indicates that the star
is earlier that previously assumed based on its optical colours: it is an
09e star. 

\subsection{Probing The Circumstellar Disk}

No far red optical spectra have been previously published for V635 Cas. 
Dramatic spectral variability occurred. The \ha line changed from emission
to absorption on a timescale of four months or less (Unger 1993). This is
the first time that a phase change has been seen in this system. The phase
change was also reflected in the Paschen and He~I lines and by the low in
the $J H K$ lightcurve. We interpreted this as a disk loss event which was
discussed by Unger (1993) and in Paper I.  Here we will interpret the
emission line data. 

With sufficient resolution many of the emission lines in Be stars are
double peaked.  Huang (1972) interpreted this in terms of a simple model
consisting of a disc rotating about the star. The outer radius of the
emission region can be estimated using the ratio of the peak separation to
the star's rotational velocity. 

The spectral resolution of our observations is not sufficient to show the
expected double peak structure of the lines. Previous \ha spectra of
higher resolution (0.8 \AA\ dispersion, 1.6 \AA\ resolution) indicate that
the peak separation remains fairly constant and that it was $\sim$462
km~s$^{-1}$ in 1991 August 28 (Unger 1993). The unresolved spectra are
consistent with either this or a narrower value. 

\subsubsection{He\,{\sc i} Lines}

The He\,{\sc i} emission is clearly seen at $\lambda$ 6678 \AA\ and
$\lambda$ 7065 \AA\ on 1991 October 16 and 27. Hence the line at $\lambda$
7281 \AA\ is probably He\,{\sc i} and not an artifact introduced by the
removal of the variable telluric H$_{2}$O absorption band.

\subsubsection{Paschen Lines}

Generally, Be stars with earlier spectral type have stronger emission in
Paschen lines (Andrillat et al. 1990).  We observe P19--P11 in emission on
1991 October 16 and 20, and weaker emission from P18--P11 on 1991 October
27. The emission lines disappear, with P13, P12 and P11 in absorption on
1992 February 19.

We hoped to be able to constrain the electron density in the disk by
investigating the relative Paschen line strengths. The line fluxes were
dereddened using a standard Galactic law, $R = A(V)/E(B - V) = 3.1$ (Rieke
\& Lebofsky 1985; Howarth 1983) and assuming $E(B - V) = 1.5$ for V635 Cas
(Hutchings \& Crampton 1981). It is difficult to fix the continuum level
between the various Paschen lines due to the broad emission wings and
blending with other lines (see Table 1). 

Line ratios were calculated for the unblended lines (i.e. P19, P17, P14,
P12, P11). We estimate that the errors in the line fluxes are $\sim$30\%.
These ratios were compared with case B optically thin recombination line
strengths for a {\it T$_{e}$} = $10^{4}$ K, {\it n$_{e}$} = $10^{8}$
cm$^{-3}$ and {\it T$_{e}$} = $10^{4}$ K, {\it n$_{e}$} = $10^{10}$
cm$^{-3}$ plasma (Hummer \& Storey 1987; Storey \& Hummer 1995). The line
ratios were also compared with the optically thick line ratios based on
the simple assumption that the disk is a {\it T} = $10^4$ K blackbody with
the same line widths and emitting areas for all the observed Paschen
lines. 

General results for Be star systems indicate that P$\beta$ and P$\gamma$
line ratios are well away from the case B values (Sellgren \& Smith 1992)
and that P19 and higher Paschen lines are consistent with optically thin
emission (Briot 1981). However, these data have a turnover in the relative
line strengths at both P17 and P11. For example on 1991 October 16 the
relative line fluxes for P17, P14, P12, P11, normalised to P17 are
1.0:2.4:1.9:1.5. This means that we cannot constrain the disk density. For
optically thin emission we would expect that the relative line strengths
increase as you descend the Paschen series, the opposite is true for
optically thick emission from a blackbody.  To constrain these models we
require observations of additional lines in the series and at a higher
resolution so that the lines are not blended.

\begin{figure*}
\rotate[l]{
\epsfxsize=100mm
\epsfbox[100 100 530 760]{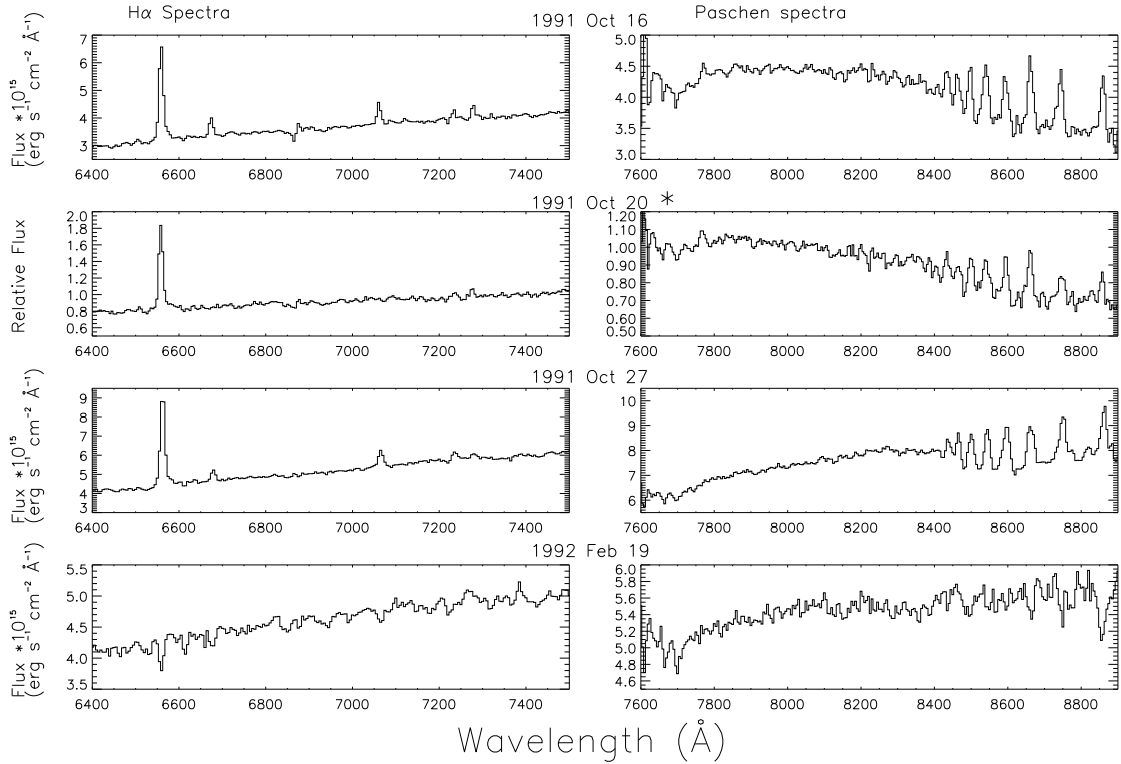}}
\caption[]{V635 Cas $\lambda\lambda$ 6400 -- 8900 \AA\ spectra}
\end{figure*}

\subsubsection{O\,{\sc i} Lines}

The O\,{\sc i} $\lambda$\ 8446 \AA\ emission is more frequent in early
type stars (Andrillat et al. 1990) and if seen is always present in
emission (Andrillat 1986). This line is blended with P18 and we would
expect it to be in emission when the O\,{\sc i} $\lambda$ 7772-74-75 \AA\
lines are in emission. The O\,{\sc i} $\lambda$ 8446 \AA\ line has a
greater tendency to go into emission than the O {\sc i} $\lambda$
7772-74-75 \AA\ line due to Bowen fluorescence (Bowen 1947).  Assuming the
lines are optically thin and adopting the P19 equivalent width for P18,
which would under estimate the strength of P18, we obtain equivalent
widths of -1.2 \AA\ and -2.0 \AA\ for O\,{\sc i} $\lambda$ 8446 \AA\ on
1991 October 16 and 20, respectively. The measured equivalent widths of of
O\,{\sc i} $\lambda$ 7772-74-75 \AA\ on 1991 October 16 and 20 are -0.5
\AA\ and -1.6 \AA, respectively.  We may have underestimated the strength
of the O\,{\sc i} $\lambda$ 8446 \AA\ line but our values are well below
the observed ratio of $\sim$4 which has been reported for these lines in
other Be stars (Jaschek et al 1993).

\subsubsection{Other Possible Features}

There are two absorption lines present in all the spectra at $\lambda$
7665 and $\lambda$ 7699 \AA\ which could be K\,{\sc i} lines.  These lines
have low excitation potentials of 1.6 eV but more importantly the
ionization potential of potassium is 4.3 eV, so any emission must be
shielded from the strong UV flux, i.e the line is probably formed in the
outer regions of the disk. Alternatively the K\,{\sc i} lines could be due
to interstellar absorption. 

There are two possible Fe\,{\sc ii} emission lines at $\sim$6480 \AA\ and
6508 \AA. The Fe\,{\sc ii} has a low excitation potential, $\sim$4--5eV,
and we may expect the line to be formed in the outer disk if we assume the
excitation potential is correlated with the disk size as in the Balmer
series. However, the Fe\,{\sc ii} line can be subject to Ly$\alpha$
fluorescence and hence we could also see it in the inner parts of the disk
(Slettebak et al 1992). Higher resolution spectra to determine the
positions of all of these lines and the individual peak separations are
needed. 

The emission line at $\lambda$ 7235 \AA\ is present in all the 1991
October spectra.  Assuming it is not an artefact due to the removal of the
telluric H$_{2}$O absorption band we tentatively identify it as C{\sc ii}
$\lambda$ 7234 \AA.  This line is excitable by resonance fluorescence from
the UV continuum and we would expect the line to be formed near the star
(Williams \& Ferguson 1983). 

There are possibly some N{\sc i} $\lambda\lambda$ 8629, 8680-83-86,
8703-12-19 \AA\ emission lines. If neutral nitrogen is seen it is always
in emission and it is more frequent in early type spectral classes
(Jaschek et al 1992;  Andrillat et al 1990). 

Finally there is a possible emission feature at $\lambda$ 8810 \AA\ on
1991 October 27.  Higher resolution spectra are needed to confirm the
presence of these lines.

\section{Conclusions}

V635 Cas is now classified as an O9e star. We have demonstrated that a
true spectral classification, although difficult to obtain, is needed to
accurately model the BeXRBs. The filling in of critical lines by the disk
emission precludes a determination of the luminosity class. We hope that
future observations during a disk-loss event will improve on this result. 

The far red spectra are an ideal probe of the circumstellar disk.  Higher
resolution spectra including additional lines in the Paschen series would
enable us to confirm various line blends/strengths and measure the peak
separation between the wings in the individual lines.  This would enable
us to model the size, density and velocity distribution in the disk. In
particular the line profiles would help us parameterize the tidal effects
of the neutron star at periastron. 

Clearly this is a complex system as demonstrated by the dramatic spectral
and photometric changes that occur on a short timescale of months.  A
precise determination of the spectral class, luminosity and $v \sin i$
requires frequent high resolution simultaneous multi-wavelength
observations in order to catch the system in a disk-less state. In
particular infrared observations of the H{\sc i} lines are needed to
further constrain the disk models.

\section*{Acknowledgements}

We thank the support astronomers at the William Herschel telescope (WHT)
for our service mode observations. The WHT is operated on La Palma by the
Royal Greenwich Observatory at the Spanish Observatorio del Roque de los
Muchachos of the Instituto de Astrofisica de Canarias. 
Michigan-Dartmouth-MIT Observatory is owned and operated by a consortium
of the University of Michigan, Dartmouth College, and the Massachusetts
Institute of Technology.  Our thanks to Diane Roussel-Dupre, for calling
our attention to the outbursts of this object.  This research made use of
the Simbad database, operated at CDS, Strasbourg, France. SJU was
supported in part by the Jet Propulsion Laboratory, California Institute
of Technology, under the sponsorship of the Astrophysics Division of
NASA's Office of Space Science and Applications.


\begin{thebibliography}{}
%\bibitem[year]{refid} entry


\bibitem[year]{refid} Andrillat Y., 1986, In:
Slettebak A. \& Snow T. P., (eds.) IAU Coll. 92, The Physics of Be Stars,
Cambridge and University and Press, p. 237

\bibitem[year]{refid} Andrillat Y., Jaschek M., Jaschek C., 1990,
A\&AS 84, 11

\bibitem[year]{refid} Bowen S. I., 1947, PASP 59, 196

\bibitem[year]{refid} Briot D., 1981, A\&A 103, 5

\bibitem[year]{refid} Carter D., Bennet C. R., Rutten R. G. M., et al., 
1993, William Herschel Telescope ISIS Users' Manual

\bibitem[year]{refid} Cominsky L., Clark G. W., Li F., Mayer W., Rappaport
S., 1978, Nat 273, 367

\bibitem[year]{refid} Filippenko A. V., 1982, PASP 94, 715

\bibitem[year]{refid} Howarth I. D., 1983, MNRAS 203, 301

\bibitem[year]{refid} Howarth I. D., Murray J., Mills D., Berry D. S.,
1996, Starlink User Note 50.19, Rutherford Appleton Laboratory

\bibitem[year]{refid} Huang S. S., 1972, ApJ 171, 549
 
\bibitem[year]{refid} Hummer D. G., Storey P. J., 1987, MNRAS 224, 801

\bibitem[year]{refid} Hutchings J. B., Crampton D., 1981, ApJ 247, 222

\bibitem[year]{refid} Jaschek M., Jaschek C., Andrillat Y., Houziaux L.,
1992, MNRAS 254, 413

\bibitem[year]{refid} Jaschek M., Jaschek C., Andrillat Y., 1993, A\&AS
97, 781

\bibitem[year]{refid} Johns M., Koski A., Canizares C., McClintock J.,
1978, I.A.U. Circ. No. 3171

\bibitem[year]{refid} Kriss G., Cominsky L., Remillard R., Williams G.,
Thorenstein J., 1983, ApJ 266, 806

\bibitem[year]{refid} Luppino G. A., 1989, PASP 101, 931

\bibitem[year]{refid} Meinel A., Aveni A., Stockton M., 1975,
Catalogue of Emission Lines In Astrophysical Objects,
University of Arizona

\bibitem[year]{refid} Mendelson H., Mazeh T., 1991, MNRAS 250, 373

\bibitem[year]{refid} Negueruela I., Grove J. E., Coe M. J. et al., 1997,
MNRAS 284, 859 (Paper I)

\bibitem[year]{refid} Oke J. B., 1974, ApJS 27, 210

\bibitem[year]{refid} Oke J. B., Gunn J. E., 1983, ApJ 266, 713

\bibitem[year]{refid} Rappaport S., Clark G. W., Cominsky L., Joss P. C.,
Li F., 1978, ApJ 224, L1

\bibitem[year]{refid} Rieke G. H., Lebofsky M. J., 1985, ApJ 288, 618


\bibitem[year]{refid} Sellgren K., Smith R. G., 1992, ApJ 388, 178

\bibitem[year]{refid} Shortridge K., Meyerdicks H., 1996, Starlink User
Note 86.11, Rutherford Appleton Laboratory

\bibitem[year]{refid}Slettebak A., Collins-II G. W., Truax R., 1992, ApJS
81, 335

\bibitem[year]{refid} Storey P. J., Hummer D. G., 1995, MNRAS 272, 41

\bibitem[year]{refid} Unger S. J., 1993, PhD Thesis, University of
Southampton

\bibitem[year]{refid} Wade R. A., Horne K., 1988, ApJ 325, 411

\bibitem[year]{refid} Walborn N. L., 1971, ApJS 23, 257

\bibitem[year]{refid} Walborn N. L., Fitzpatrick E. L., 1990, PASP 102, 379

\bibitem[year]{refid} Williams R. E., Ferguson D. H., 1983, In: Livio L.,
Shaviv G. (eds.), IAU Coll. 72, Cataclysmic Variables and Related Objects,
Reidel, p. 97

 
\end{thebibliography}
\end{document}